\documentclass{aastex631}
\usepackage{natbib}
\usepackage{xspace}
\usepackage{enumitem}

\shorttitle{Long-Duration Nonthermal Plasma Motions}
\shortauthors{Gou \& Reeves}

\accepted{May 23, 2025}
\submitjournal{ApJ}

\begin{document}

\title{Long-Duration Nonthermal Motions in the Supra-Arcade and \\
       Loop-Top Region During an Eruptive Solar Flare}

\author[0000-0003-0510-3175]{Tingyu Gou}
\author[0000-0002-6903-6832]{Katharine K. Reeves}
\affiliation{Center for Astrophysics $|$ Harvard $\&$ Smithsonian, 60 Garden Street, Cambridge, MA 02138, USA \\ \url{tingyu.gou@cfa.harvard.edu}}

\begin{abstract}
Solar flares are widely accepted to be powered by magnetic reconnection that involves complex dynamics in various scales. The flare supra-arcade and loop-top region, directly impacted by fast reconnection downflows, contains a wealth of microscopic dynamics, which are, however, difficult to resolve in imaging. We present simultaneous spectroscopic and imaging observations of hot flaring plasma above the loop tops by IRIS and SDO/AIA. IRIS continuously observed high-temperature \ion{Fe}{21} 1354.08~\AA\ spectral emissions throughout the long-duration gradual phase of the X-class flare. We found weak Doppler blue shifts near the loop-top region, indicative of bulk plasma motions from chromospheric evaporation based on the 3D flare loop orientation. Strong nonthermal velocities are detected at the bottom of the flare supra-arcade fan/plasma sheet, suggestive of the presence of turbulence in the flare current sheet region. In addition, disorganized nonthermal plasma motions are constantly detected until the very end of the flare, indicating irregular unresolved plasma flows in the cusp and loop-top region. The spatial and temporal evolution of spectral parameters follow the dynamics resulting from on-going magnetic reconnection during the prolonged gradual phase. The long-lasting nonthermal plasma motions may contribute to the high and steady temperature of flaring plasmas above flare loops.
\end{abstract}

\keywords{Solar flares (1496); Solar extreme ultraviolet emission (1493); Solar flare spectra (1982); Solar magnetic reconnection (1504); Active solar corona (1988)}

\section{Introduction} \label{sec:intro}

Solar flares are magnificent phenomena responsible for the sudden and rapid release of magnetic energy in the solar atmosphere. The energy release process is widely accepted to be powered by magnetic reconnection, which converts the stored magnetic free energy into thermal and nonthermal kinetic energies of hot plasmas and energetic particles in the flare, as well as bulk kinetic energy of the coronal mass ejection (CME) during the eruption. As a result, flaring plasmas are intensively heated to above 10~MK, accompanying with a wealth of dynamics in the reconnection process. Although a standard picture of flares has been built for years \citep[referred to as the `CSHKP' model following][]{Carmichael1964,Sturrock1968,Hirayama1974,Kopp1976}, the detailed dynamics associated with magnetic reconnection and the energy release process is not fully understood.

High-cadence, high-resolution observation and numerical modeling have demonstrated the presence of plentiful dynamics occurring during flares, especially in the region above flare loops where magnetic reconnection is believed to occur. Cool inflows move perpendicularly into the reconnection current sheet region and hot outflows move along it either upward or downward to the flare loops \citep[e.g.,][]{Savage2012,LiuR2013,Gou2017,Longcope2018}. Particularly a lot of downward-moving structures appear in the above-loop-top region, in the form of super-arcade downflows/downflowing loops \citep[SADs/SADLs; e.g.,][]{McKenzie1999,Innes2003,Savage2012,Reeves2017,Gou2020} and shrinking cusp-shaped loops \citep{Forbes1996,Reeves2008,LiuW2013}, which are thought to be direct consequences of magnetic reconnection. When the fast-descending downflows fall close to dense flare loop tops, a termination shock is expected to form, which can accelerate particles to high energies \citep{Forbes1986,Takasao2015,ChenB2015}. A pair of slow-mode shocks is also predicted to form bifurcating the current sheet in the cusp region, which can greatly heat plasmas \citep{Petschek1964,Priest1986,Tsuneta1996,Gou2015}. The plentiful dynamics occurring in the super-arcade and loop-top region play a crucial role in understanding the thermal and nonthermal processes in flares.

The current sheet above flare loops is an essential feature in the flare reconnection. In imaging observations, plasmas surrounding a reconnection current sheet appear as either a linear plasma sheet or a broad plasma fan, corresponding to an edge-on or face-on view for the observer \citep[e.g.,][]{Ciaravella2008,Innes2014,Hanneman2014,Seaton2017,ChenB2020,Gou2024}. Thermal diagnostics on the plasma sheet usually show high temperatures, particularly associated with hot almost isothermal plasmas in a steady state \citep{Gou2024}. While the analysis of high-resolution EUV imaging suggests a possible presence of turbulence in the current sheet region \citep{Xie2024,LiuR2021,Cheng2018}, which is predicted and well reproduced in numerical simulations \citep[e.g.,][]{Barta2011,Mei2012,Huang2016,Shen2022,Wang2023}. The current sheet during fast reconnection is expected to be highly fragmental consisting of magnetic islands (also termed plasmoids) in different scales due to the tearing mode and coalescence instabilities \citep{Furth1963}, and observations have provided evidence for the tearing of flare current sheet into multiple plasmoids that move along the sheet and coalesce into larger ones \citep{Gou2019}. However, there still exists a significant gap between the observational, macro-scale dynamics and the real, micro-scale energy release, and remote-sensing imaging is extremely limited for investigating the small-scale dynamics during fast flare reconnection. Other approaches are needed to understand the detailed physical process.

Spectroscopic observations provide a unique diagnosis on plasma motions and potential turbulent behaviors during flares, with the ability to resolve plasma flows and microscopic dynamics unrevealed by direct imaging. Details of dynamics in flares have been found, benefiting from high-cadence and high-resolution spectral observations such as those obtained by the Interface Region Imaging Spectrograph \citep[IRIS;][]{DePontieu2014}. For example, completely red-shifted emissions in the high-temperature spectral line indicate hot, downward-moving outflows \citep{Tian2014}; large Doppler shifts co-spatial with loop-top hard X-ray (HXR) emissions are possible deflection flows associated with a flare termination shock \citep{Polito2018}; periodic oscillations in Doppler velocities are associated with a `magnetic tuning fork' created by reconnection outflows colliding with flare loop tops \citep{Reeves2020}; nonthermal broadening in the spectral line implying plasma turbulence in the loop-top region could play a role in particle acceleration through turbulent energy dissipation \citep{Ashfield2024}. 

Direct spectral observations of the flare plasma sheet are rare. A well-known case is the observation of the 2017 September 10 X8.2 flare by the EUV Imaging Spectrometer \citep[EIS;][]{Culhane2007} onboard Hinode \citep{Kosugi2007}, in which excess line broadening is detected in the plasma sheet as indications for turbulent magnetic reconnection \citep{Warren2018,LiY2018,French2020}. IRIS observed the \ion{Fe}{21} spectral emissions in the current sheet and cusp region of an M9.2 flare on 2015 March 7, where the nonthermal line broadening is comparable to that from turbulent plasma flows in the reconnection region in MHD simulation \citep{Shen2023}. Nonthermal motions above the flare loop tops are also detected in a few flares by Hinode/EIS \citep[e.g.,][]{Doschek2014}, which may be associated with plasma turbulence. Due to limited favorable observations, spectroscopic diagnostics on detailed dynamics and nonthermal processes in flares remain largely unexplored.

We study an eruptive X-class flare that has ideal spectroscopic observations by IRIS. The Atmospheric Imaging Assembly \citep[AIA;][]{Lemen2012} onboard the Solar Dynamics Observatory \citep[SDO;][]{Pesnell2012} provides simultaneous multi-wavelength imaging for this long-duration event. Particularly, IRIS performed continuous observation covering the complete flare, providing spectroscopic insights for the flare dynamics during its whole life. We found large nonthermal velocities in the flare supra-arcade and loop-top region, which are present throughout the prolonged flare gradual phase and vary in both time and space. We present the observations and analyses of data in Section~\ref{sec:obs}, and discuss the findings in Section~\ref{sec:dis}. A brief summary and conclusion is given in Section~\ref{sec:summary}.

\section{Observation and Analysis} \label{sec:obs}

\begin{figure}[htbp]
	\centering
	\includegraphics[width=\textwidth]{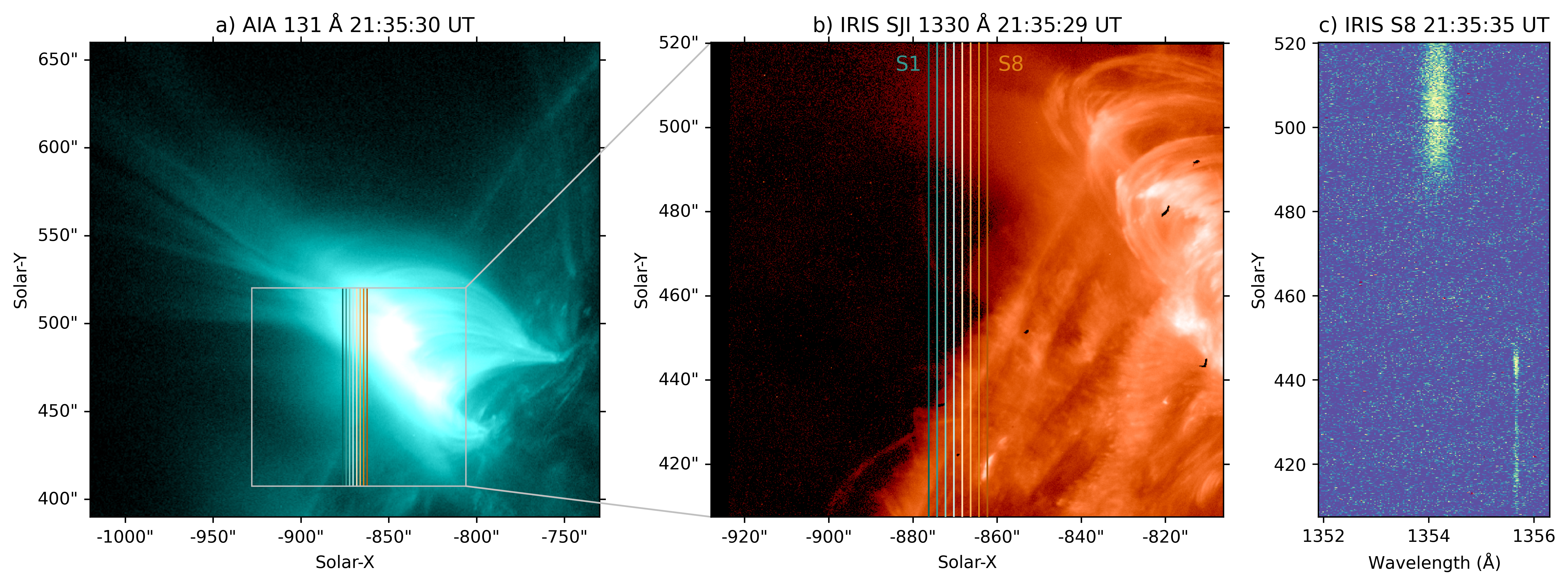}
	\caption{\small SDO/AIA and IRIS observation of the flare. From left to right are AIA 131~\AA\ image, IRIS SJI 1330~\AA\ image, and IRIS spectrum in the \ion{O}{1} 1354~\AA\ spectral window. The gray rectangle in panel (a) indicates the FOV of IRIS SJI images. The vertical colored lines in panels (a,b) show locations of IRIS slits S1 -- S8. The spectrum at S8 is shown in panel (c).}
	\label{fig:obs}
\end{figure}

We investigate an eruptive X2.3 flare occurring on 2023 February 17, in the NOAA active region (AR) 13229 near the northeast limb (N25E67). The flare starts at 19:38~UT and peaks at 20:16~UT in the GOES soft X-ray (SXR) flux. AIA observed the event as a typical two-ribbon flare, showing two quasi-parallel ribbons connected by cusp-shaped flare loops in the wake of an eruption. The impulsive eruption is followed by a long-duration gradual decay of the flare, which lasts over 4 hours. Due to a slightly oblique view angle, the plasmas in the supra-arcade region include both edge-on and face-on characteristics, showing elongated plasma spikes, a spread plasma sheet/fan, and persistent dynamic downflows during the long-duration flare gradual phase (Fig.~\ref{fig:obs}a). 

IRIS observed this event with a continuous 5-hour program, covering both the impulsive phase and the long-lasting gradual phase of the flare. IRIS captured the southeast part of post-flare loops in slit jaw images (SJIs), which were taken in the 1330~\AA\ channel in a $119''\times 119''$ field of view (FOV) with a 10~s cadence (Fig.~\ref{fig:obs}b). IRIS spectra were taken in an 8-step raster with a spatial step of 2$''$ and a temporal cadence of 9.6~s per step, and it provided a high spatial scale of 1$''$/6 and a spectral resolution of about 0.026~\AA\ in the FUV band. We refer to the IRIS slit locations as S1 -- S8 from the solar east to west in this study. Hot emissions from the spectral line \ion{Fe}{21} 1354.08~\AA\ (with a formation temperature of $\log T$ = 7.05, $\sim$11~MK) are observed near the flare loop-top region above the solar limb (Fig.\ref{fig:obs}c), which are also detected in SJI 1330~\AA\ images as diffusive, faint plasmas above cooler flare loops in the \ion{C}{2} 1334/1335~\AA\ lines (Fig.\ref{fig:obs}b). We focus on this high-temperature \ion{Fe}{21} spectral line and the flare loop-top region and study their spatial and temporal variations during the prolonged flare.

\subsection{Spectroscopic Diagnostics from IRIS}

\begin{figure}[htbp]
	\centering
	\includegraphics[width=\textwidth]{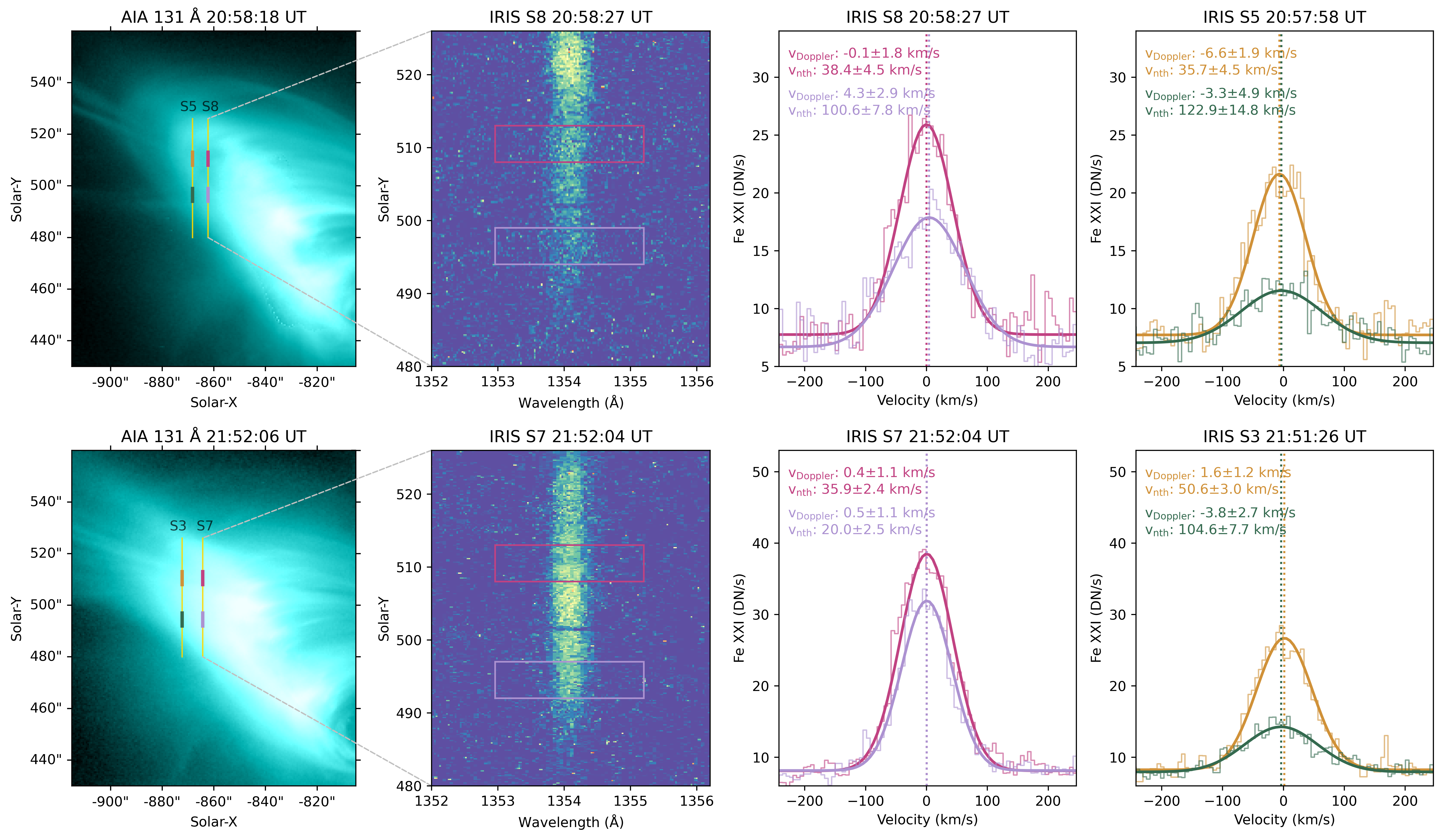}
	\caption{\small IRIS \ion{Fe}{21} line spectra at some example locations in the flare loop-top region. Top and bottom panels show the observations from two raster scans at around 20:58~UT and 21:52~UT, respectively. From left to right shows AIA 131~\AA\ images, IRIS spectra, and \ion{Fe}{21} spectral profiles sampled from two slits. The spectra from two solar-y positions along each slit are plotted in the same panel, indicated by two vertical bars with the same colors in the AIA image. Each sample location is integrated over a solar-y range of 5'' (30 IRIS pixels; within the height of colored rectangles in the spectrum) in order to obtain enough counts. The width of colored rectangles in the spectrum indicate the wavelength range used for spectral fitting. The spectral profile is fitted with a Gaussian function, and the obtained Doppler and nonthermal velocities are labeled in the same colors.}
	\label{fig:spectra}
\end{figure}

\begin{figure}[htbp]
	\centering
	\includegraphics[width=\textwidth]{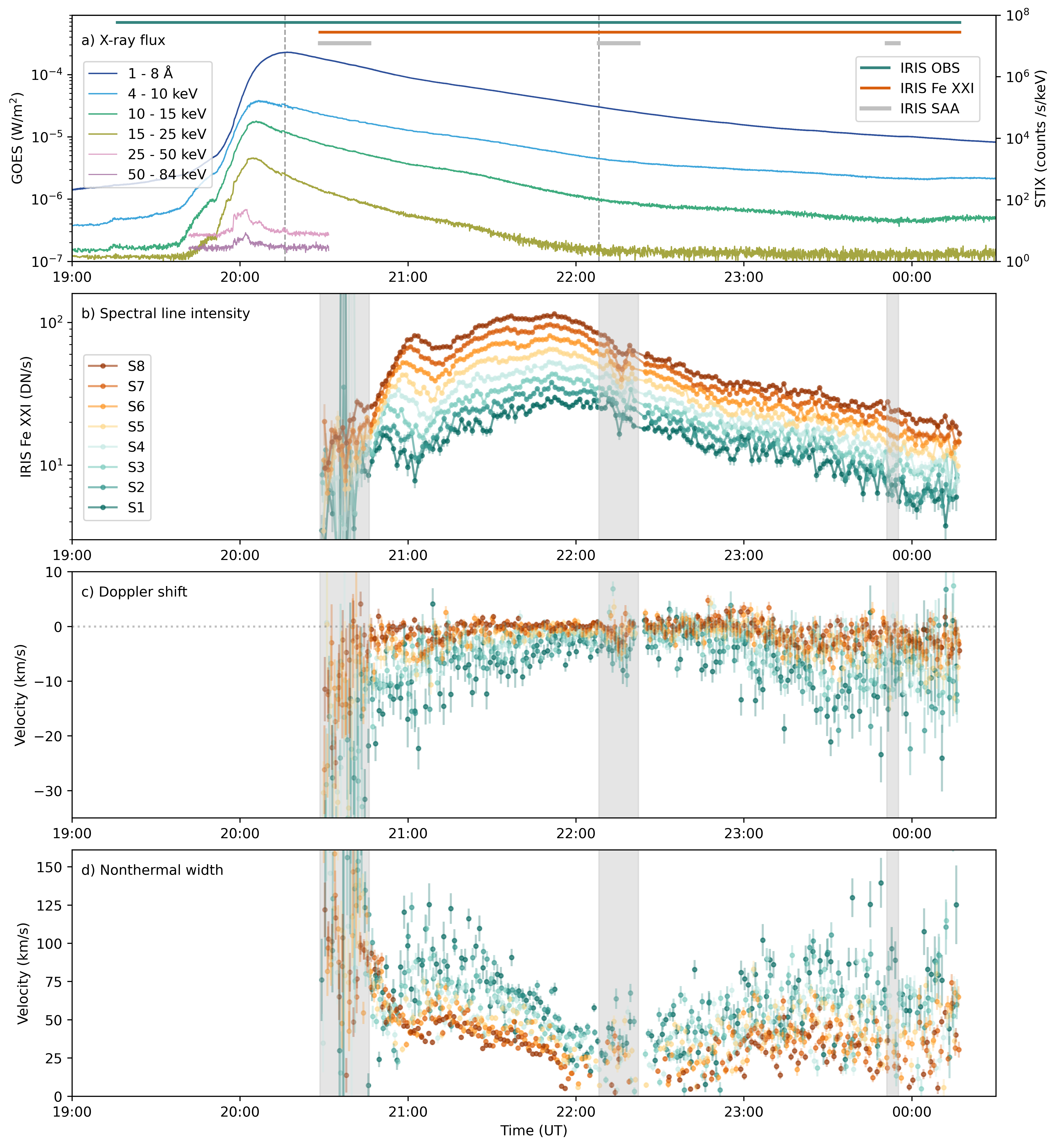}
	\caption{\small Temporal evolution of X-ray flux and spectral parameters during the flare. a) GOES SXR and STIX HXR fluxes. The colored horizontal lines in the top mark the time periods of IRIS observation, detection of \ion{Fe}{21} spectral emission, and South Atlantic Anomaly (SAA) passes of IRIS orbit during the flare, respectively. The two vertical dashed lines mark the flare SXR peak and the separation of EGP and LGP. b--d) Spectral line intensity, Doppler and nonthermal velocities obtained from spectral fittings in slit locations S1 -- S8, shown in different colors. The gray shaded regions indicate the time periods when IRIS passes the SAA, where the results can be safely ignored due to strong impacts of energetic particles.}
	\label{fig:timeline}
\end{figure}

We examine the spectral line emissions and profiles observed by IRIS during the flare. Relatively weak emissions from the \ion{Fe}{21} line were captured from the off-limb flaring plasma. To increase the signal-to-noise ratio, we perform wavelet denoising on the IRIS spectral images using the ‘db1’ wavelet and ‘BayesShrink’ thresholding algorithm. The bottom of IRIS slits are located on the solar disk and captured cool emissions from the \ion{O}{1} 1355.598~\AA\ line in the same spectral window of \ion{Fe}{21} (Fig.~\ref{fig:obs}c). Thus we use \ion{O}{1} line for wavelength calibration to correct the orbital variation and obtain the rest wavelength of the spectral line. The optically-thin \ion{Fe}{21} line shows a typical Gaussian-shaped profile during the whole flare process (Fig.~\ref{fig:spectra}), demonstrating contributions from isothermal coronal plasmas. We fit a Gaussian to the spectrum and obtain the total line intensity, Doppler shift, and line broadening. By deducting the thermal broadening of the spectral line ($w_{th}\approx$ 0.43~\AA\ taking a formation temperature of $\log T$ = 7.05) and the instrumental broadening ($w_{I}$ = 26~m\AA\ in the IRIS FUV band), we obtain the nonthermal width, $w_{nth} = \sqrt{w^2_{FWHM} - w^2_{th} - w^2_I}$. 

We show the IRIS \ion{Fe}{21} spectra and fitting results from some representative locations in Fig.~\ref{fig:spectra}. The flare plasmas show small Doppler shifts of a few km/s. However, significant nonthermal velocities are detected, varying from tens to $>$100 km/s in different locations. The nonthermal velocity is measured as high as $>$120~km/s in the supra-arcade region (e.g., the spectrum in dark green at 20:58~UT in Fig.~\ref{fig:spectra}), where diffusive plasmas are also observed in the AIA 131~\AA\ image that is dominated by the \ion{Fe}{21} contribution. As the flare proceeds, spectral emissions and characteristics change. We examine the spectral parameters in detail at all slit positions. 

Considering the observed \ion{Fe}{21} emissions are tenuous (especially during the late flare phase) and the spectral line profiles along the slit show a similar Gaussian shape (Fig.~\ref{fig:spectra}), we integrate the spectral emissions from the entire loop-top region for each slit position (y = [480'', 526'']; we refer IRIS S1 -- S8 in this solar-y range hereafter). This approach ensures sufficient signal to increase the accuracy of spectral line fitting, although it ignores the spatial resolution along the slit since the loop-top region is relatively narrow in the observation. As a compensation, IRIS slits S1 -- S8 are located in different projection heights in the plane of sky, thus enabling us to resolve the spatial distribution with height above flare loops (i.e., slits on the left cover larger altitudes). We perform Gaussian fits to the spectral line profiles in all slit positions and at different times during the flare. The fitted spectral parameters in relation to flare X-ray emissions are shown in Fig.~\ref{fig:timeline}.

IRIS focused on the flare region and started its observation at 19:16~UT before the flare onset. \ion{Fe}{21} spectral line emissions are detected at $\sim$20:28~UT after the flare SXR peak (Fig.~\ref{fig:timeline}; during an SAA pass of the IRIS orbit), when the growing post-flare arcade shows up under the IRIS slits. Spectral emissions are continuously observed under the slits during the long-duration gradual phase until IRIS stopped the observing program at 00:17~UT. For the $\sim$4-hour \ion{Fe}{21} spectral emissions during the flare gradual phase (Fig.~\ref{fig:timeline}b-d), we obtain the following results.
\begin{enumerate}[noitemsep,topsep=-\parskip]
    \item The spectral line intensities show similar behaviors in different slit positions, except that they are stronger in slits at lower altitudes (in orange) and reach the peaks successively with time in different heights. This result demonstrates the capability of investigating plasmas at different altitudes in the flare.
    \item The spectral line profiles are generally blue-shifted, indicative of bulk plasma motions with light-of-sight velocities of $\leq$ 20~km/s.
    \item Significant nonthermal line broadenings are detected throughout the flare gradual phase, with velocities up to $\sim$130~km/s.
    \item For different slit positions, those at higher altitudes (in green) generally have stronger Doppler shifts and much larger nonthermal velocities than those at lower altitudes (in orange). 
    \item During the prolonged flare gradual phase, the spectral line intensity exhibits an initial increase and then a relatively smooth decrease, while the Doppler velocity and nonthermal width show a distinctly different evolution, both of which decrease first and increase again with time.
    \item The spectral parameters, including the line intensity, Doppler shift and nonthermal width, exhibit a relatively organized and smoothed temporal evolution in the first period of gradual phase (e.g., before 22~UT), while show a more disordered variation toward the flare end.
\end{enumerate}

Based on the temporal evolution of spectral parameters, we divide the long-duration decay of the flare into two distinct episodes, referred to as early gradual phase (EGP) and late gradual phase (LGP). It is difficult to define the exact time separating EGP and LGP due to an SAA pass that affects the spectral observation (Fig.~\ref{fig:timeline}). We generally separate the two phases by the start of the second SAA at $\sim$22:08~UT, considering the line intensities from all slits start to decline significantly afterward. The different temporal evolution and spatial variation of spectral parameters indicate distinct characteristics of hot flaring plasmas near the loop-top region. We investigate the spectral characteristics in combination with AIA EUV imaging.

\subsection{Flare Dynamics in AIA}

\begin{figure}[htbp]
	\centering
	\includegraphics[width=\textwidth]{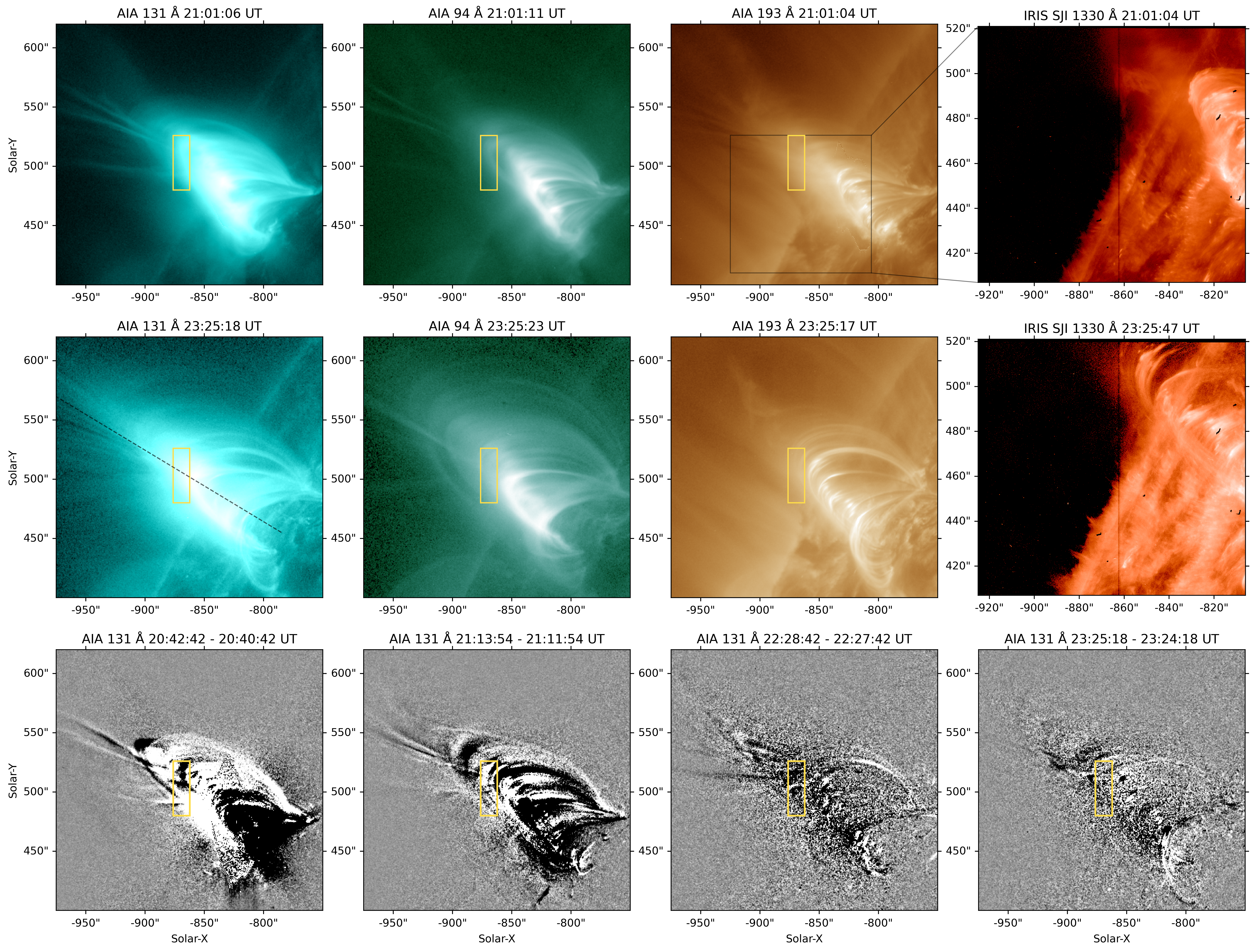}
	\caption{\small Multi-wavelength observation of post-flare loops. Top and middle panels show AIA 131~\AA, 94~\AA, 193~\AA, and IRIS SJI 1330~\AA\ images during the EGP and LGP, respectively. Bottom panels show AIA 131~\AA\ running difference images at four different time instants. The yellow box shows the region of IRIS slit coverage where \ion{Fe}{21} spectral emissions are observed. The black rectangle in the AIA 193~\AA\ image indicated the IRIS SJI FOV. The black dashed line in the AIA 131~\AA\ image indicates the location of a virtual slit used to generate the stack plots in Fig.~\ref{fig:slit}.
    An animation of this figure shows the AIA 131~\AA\ original and running difference images from 20:45~UT to 00:20~UT, which highlights the plasma structures and dynamic evolution above flare loops during the gradual phase. The region of IRIS spectral emission is indicated as a yellow box in the animation for comparison.}
	\label{fig:maps}
\end{figure}

\begin{figure}[htbp]
	\centering
	\includegraphics[width=\textwidth]{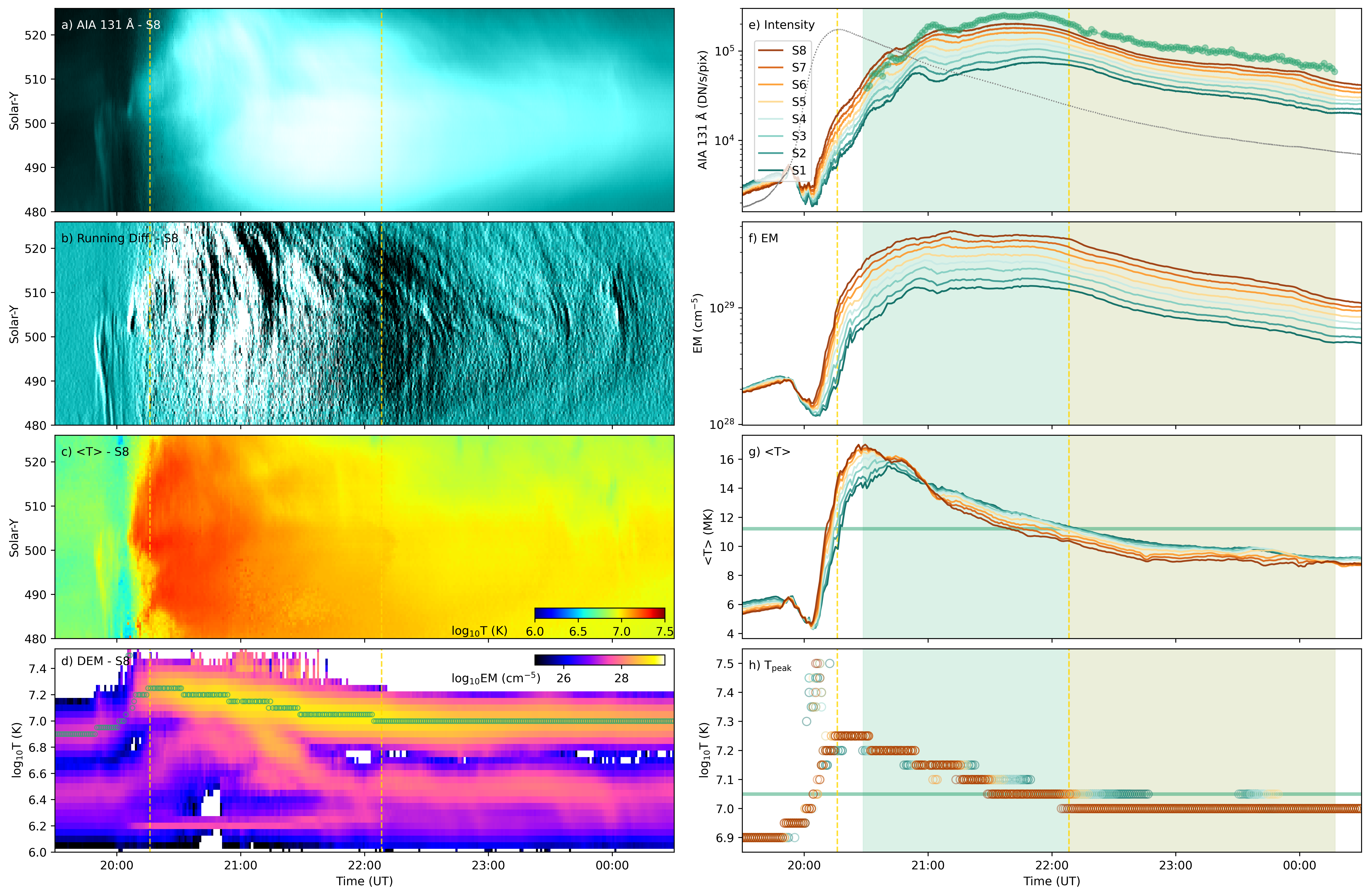}
	\caption{\small Temporal evolution of AIA intensity and DEM in the flare loop-top region. a--c) AIA 131~\AA\ intensity, its running difference, and the DEM-weighted mean temperature at the location of IRIS slit S8 in the loop-top region. d) DEM distribution at S8, averaged over the loop-top region shown in panels (a--c). The green circular symbols indicate the peak temperature of the hot DEM component. e--g) AIA 131~\AA\ intensity, total EM, and mean temperature at S1 -- S8, averaged over the loop-top region. The gray dotted curve in panel (e) shows the GOES 1--8~\AA\ flux. The green dots show the IRIS \ion{Fe}{21} line intensity at S8. h) Peak temperature of the hot DEM component at S1 -- S8. The horizontal thick lines in panels (g,h) indicate the formation temperature of \ion{Fe}{21} emission line. The colored shaded regions show the flare EGP and LGP from IRIS spectral observations, respectively. The two vertical yellow lines mark the flare peak and the separation of EGP and LGP, respectively.}
	\label{fig:dem}
\end{figure}

\begin{figure}[htbp]
	\centering
	\includegraphics[width=0.7\textwidth]{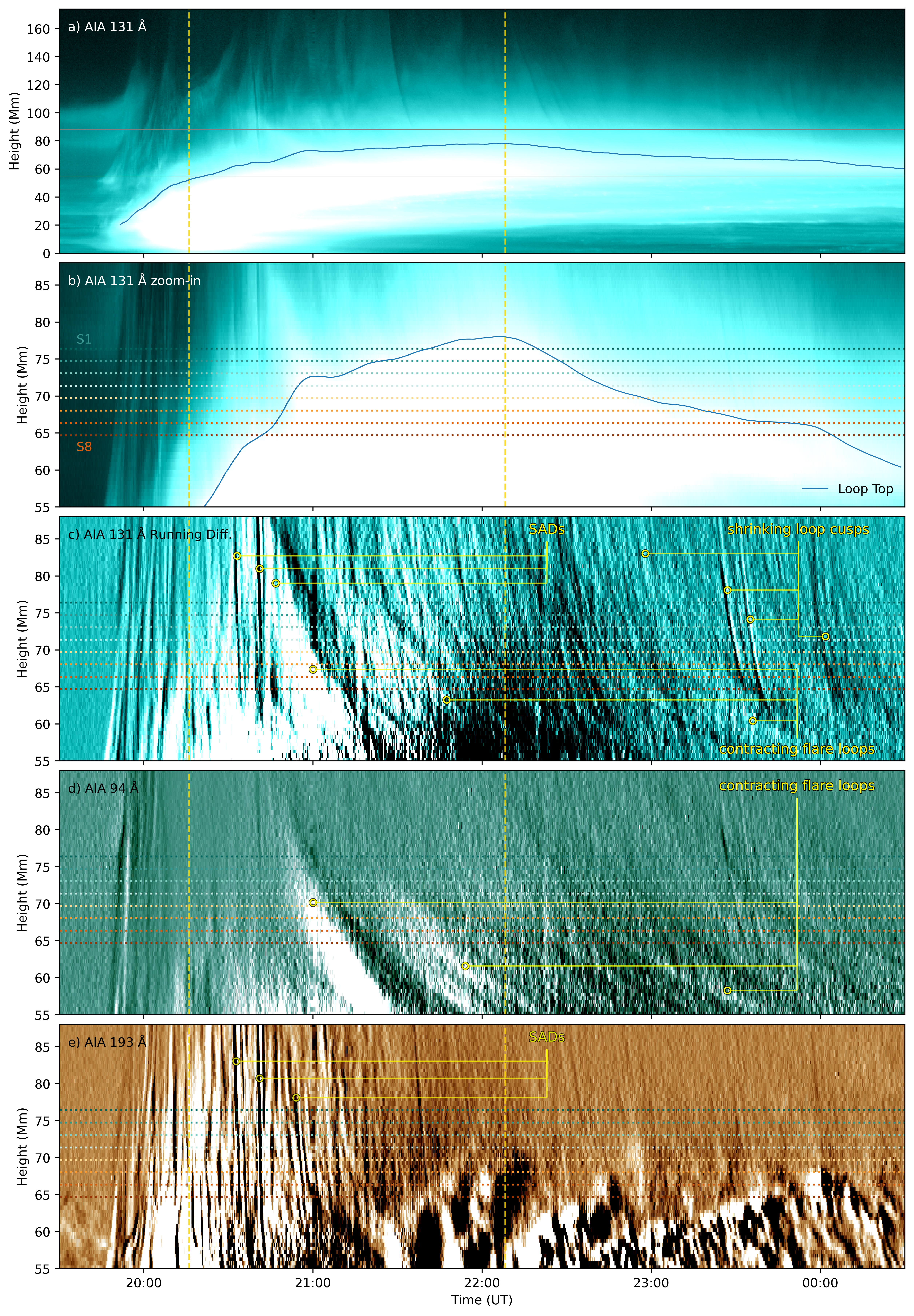}
	\caption{\small Dynamic evolution of flare loops in AIA. a) Height-time plot of flare loops in AIA 131~\AA, as seen through a virtual slit indicated by the dotted line in Fig.~\ref{fig:maps}. The blue curve show the identified height of flare loop tops in AIA 131~\AA. b--e) Height-time plots zooming into the flare loop-top region (between the two horizontal lines in panel a), made from AIA 131~\AA\ images, running difference images of AIA 131~\AA, 94~\AA, and 193~\AA, respectively. The blue curve in panel (b) shows the loop-top height taken from panel (a). The colored horizontal lines show the intersecting heights with IRIS slits S1 -- S8. Some examples of the observational features are labeled in yellow in panels (c--e). The two vertical dashed lines mark different flare phases as those in Fig.~\ref{fig:dem}.}
	\label{fig:slit}
\end{figure}

In order to compare with the IRIS high-temperature spectral observation, we investigate the flare loop-top region in AIA hot channels including 131~\AA, 94~\AA, and 193~\AA, which contain hot emissions from \ion{Fe}{21}, \ion{Fe}{18} ($\log T$ = 6.85), and \ion{Fe}{24} ($\log T$ = 7.25) during flares, respectively. AIA takes full-Sun images every 12~s in a spatial scale of 0.''6. During this intense flare, alternating long-short exposure times are applied to accommodate bright emissions from flare loops. We make composite images combining long and short exposure time pairs to remove the saturation in long-exposure images. AIA imaged various features and dynamics in the supra-arcade and loop-top region during the flare gradual phase, including for example, descending SADs, swing plasma spikes, contracting post-reconnection loops, and growing post-flare arcade (Fig.~\ref{fig:maps} and the accompanying movie). These features and dynamic evolution are observed throughout the flare gradual phase, indicating magnetic reconnection is still on-going far beyond the impulsive eruption. 

The IRIS raster captured various features under slits (see Fig.~\ref{fig:maps} and the accompanying movie), resulting from two aspects: a) the spread post-flare loops have different projected heights in the plane of sky, thus multiple optically-thin features emitting the \ion{Fe}{21} line are captured along the light of sight, for example, including both supra-arcade structures and loop tops; b) the entire flare loop system grows larger as the eruption proceeds, thus IRIS slits scan over different structures during different flare phases. The plasma structures and dynamic evolution are highlighted in AIA 131~\AA\ running difference images in Fig.~\ref{fig:maps} and the accompanying movie. We can clearly see that the IRIS raster sits on the bottom of multiple plasma spikes in the supra-arcade fan and the overlapped flare loop tops during EGP, while mostly on loop cusps and contracting flare loops during LGP. The different features captured during different phases may result in distinct evolution of spectral parameters.

We examine the AIA 131~\AA\ emissions in the loop-top region from the same locations with IRIS slits. The averaged AIA intensities along S1 -- S8 are shown in Fig.~\ref{fig:dem}(e) and the stack plot from S8 is shown in Fig.~\ref{fig:dem}(a,b) as an example. The AIA 131~\AA\ intensities exhibit similar evolution to the IRIS spectral line because of the domination of \ion{Fe}{21} contributions in this channel, and AIA images can also resolve the detailed flare loop dynamics. The 131~\AA\ emission becomes stronger after the flare peak then faint toward the flare end, and the strongest intensities are observed at y$\simeq$[485'', 505''] in the flare loop tops from about 1 hour after the flare peak. At about 21:00~UT, a bunch of loops are observed to move downward (Fig.~\ref{fig:dem}a,b), which causes successive drops of both AIA and IRIS spectral intensities from S1 to S8 (Figs.~\ref{fig:timeline}b,\ref{fig:dem}e). The contracting loops are also evident in the AIA 94~\AA\ image (Fig.~\ref{fig:maps}). The 131~\AA\ intensity becomes significantly dimmer at around 22:00~UT in the running difference images (Fig.~\ref{fig:dem}b), which coincides with the decrease of spectral intensities during LGP.

To examine the dynamic evolution of plasma structures above flare loops, we make AIA height-time plots through a virtual slit along the erupting direction (see the dashed line in Fig.~\ref{fig:maps}, using the center of the two footpoints of the flare loops as the start). We note that the IRIS SJI 1330~\AA\ channel also imaged faint \ion{Fe}{21} emissions above cool flare loops (Fig.~\ref{fig:maps}), but these emissions are too weak and diffusive to resolve detailed dynamic evolution. In AIA, a lot of downward-moving features are captured during the flare gradual phase, and different features are generally seen in different channels (Fig.~\ref{fig:slit}). We zoom into the flare loop-top region where the IRIS slits are located and find their heights (see the colored dotted lines in Fig.~\ref{fig:slit}b--e). Dynamic plasma features that cross the IRIS slits include: a) fast-descending SADs, visible in 131 and 193~\AA\ shortly after the flare SXR peak; b) contracting round flare loops, most visible in 94~\AA\ and partially in 131~\AA\ throughout the long-duration gradual phase; c) shrinking loop cusps, evident in 131~\AA, showing significantly higher velocities than contracting round flare loops. 

The distinct dynamics in the AIA emission indicate different peak-temperature responses for the supra-arcade structures and flare loops. However, all features are visible in 131~\AA, suggestive of significant contributions from the \ion{Fe}{21} line. This result demonstrates the capability of IRIS raster to capture all these plasma features. 
At around 22:00~UT, the flare loops become dark as seen in AIA 131~\AA\ difference images (Fig.~\ref{fig:slit}c). The loop dimming does not appear in the other two channels, indicative of dominant plasma emissions from \ion{Fe}{21}. The contraction of flare loops and relaxation of loop cusps can be observed until the very end, as indications of on-going magnetic reconnection.

We note that the supra-arcade plasma structures and flare loops are overlapped together along the line of sight, and their heights change as the flare proceeds. We identify the flare loop tops by selecting a threshold of the loop brightness in AIA 131~\AA, and plot the height-time evolution in Fig.~\ref{fig:slit}(a,b). The temporal evolution of loop-top height resembles that of the IRIS spectral line and AIA 131~\AA\ intensities, demonstrating the effects of loop height evolution on \ion{Fe}{21} emissions. The loop top rises rapidly after the flare SXR peak until it reaches a maximum height at about 22:07~UT, which shows a continuous growth of post-flare loops as the eruption proceeds. The identified height then decreases during LGP, which is a result of the consistent fading of flare loops several hours after the flare peak. Comparing the loop-top height with IRIS slit locations, it shows that the features captured by IRIS change a lot during the gradual phase: supra-arcade plasmas under slits in the beginning, mostly flare loop tops in the middle, more supra-arcade plasmas again toward the flare end. For different slit positions, those at high altitudes always capture more supra-arcade features than at low altitudes. 

Thus, distinct plasma features and dynamics are captured in IRIS spectra during the prolonged flare gradual phase, including the bottom of the supra-arcade fan/plasma sheet during EGP, numerous loop cusps during LGP, and superposed flare loop tops. The temporal evolution of flare loop dynamics coincides well with the spectral properties. The observations evidence strong nonthermal motions in the supra-arcade plasmas, and the nonthermal velocities are invariably larger at higher altitudes above post-flare loops.

\clearpage
\subsection{Thermal Diagnostics via DEM}

\begin{figure}[b]
	\centering
	\includegraphics[width=\textwidth]{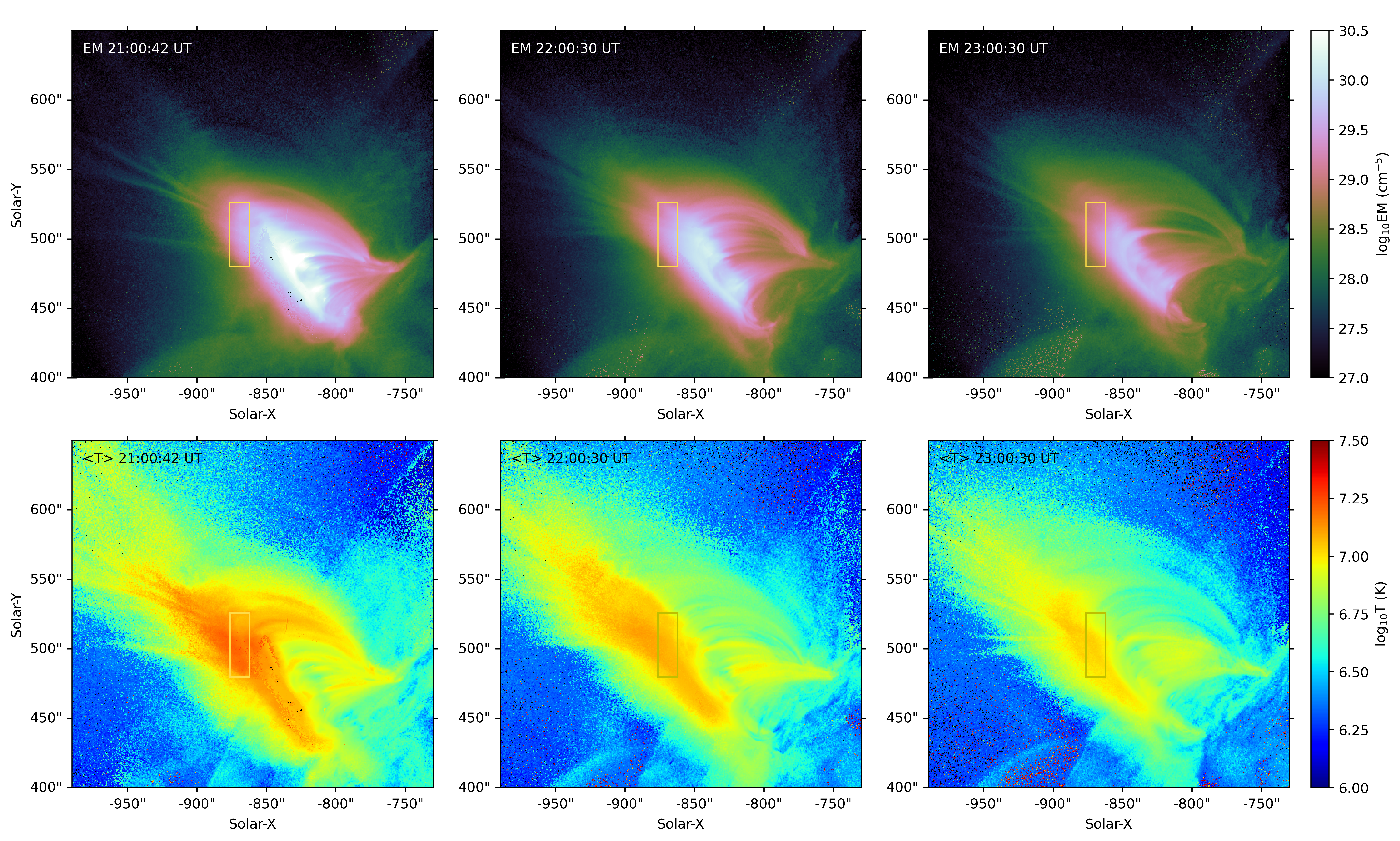}
	\caption{\small EM and mean temperature maps during the flare gradual phase. The yellow box shows the region where \ion{Fe}{21} spectral emissions are observed by IRIS.}
	\label{fig:emt}
\end{figure}

Considering the AIA and spectral intensities are results of joint contributions from the plasma temperature and density, we perform DEM analysis to investigate the thermal properties of flaring plasmas. We adopt the sparse inversion DEM method \citep{Cheung2015} using AIA data from 6 optically-thin EUV channels. DEMs are calculated in the temperature range of $\log T$ = 5.5 -- 7.6 with an interval of $\Delta\log T$ = 0.05. We obtain the total EMs and the EM-weighted mean temperature, $\langle T\rangle=\frac{\sum_{i} \mathrm{EM}(T_i) \times T_i}{\sum_{i} \mathrm{EM}(T_i)}$. The EM and mean temperature maps at three example time instants are shown in Fig.~\ref{fig:emt}. The region above round, dense flare loops always contains plasmas of the highest temperature, and the temperature there stays high ($\geq$10~MK) even several hours after the flare peak. We note that in this flare, IRIS captured spectral emissions right in the highest-temperature region (see the yellow box in Fig.~\ref{fig:emt}), which provides perfect spectroscopic insights in addition to the thermodynamic evolution.

We examine the temporal evolution of EM and $\langle T\rangle$ in the region where IRIS slit are located (Fig.~\ref{fig:dem}). The results show that IRIS detected \ion{Fe}{21} emissions only after the EM rises high enough despite the plasma temperature reaches $\log T$ = 7.05 earlier before, indicative of joint contributions from both. During the flare gradual phase, the total EM and mean temperature exhibit different evolutionary trends (Fig.~\ref{fig:dem}f,g): the EM keeps almost constant during EGP and decreases during LGP, while the temperature exhibits an intensive decrease during EGP and very small changes during LGP.

We further investigate the DEM distribution over temperature and its evolution. The DEMs at S8 are shown in Fig.~\ref{fig:dem}(d) as an example. The plasmas mainly contain two DEM components (here we only focus on hot flaring plasmas of $\log T > 6$), and the temperature ranges of the two components vary with time during the flare. Around the flare SXR peak, the hot DEM component appears in $\log T$ = 7.0 -- 7.5, with a peak at $\log T$ = 7.25; the other warm component shows increasing temperatures. During EGP, the temperatures of both DEM components drop down significantly, and especially a lot of EM moves from the hot component to the warm one, indicative of intensive cooling. During LGP, the hot DEM component remains steady in $\log T$ = 6.7 -- 7.2 with a constant peak of $\log T$ = 7.0 for over two hours, and the warm component remains in $\log T$ = 6.2 -- 6.7. During the flare gradual phase, the hot DEM component dominates over the other component in the loop-top region, thus it gives a similar peak temperature ($T_{\rm peak}$) to the EM-weighted mean temperature (Fig.~\ref{fig:dem}g,h), both of which decrease significantly at first but remain steady afterward. 

The DEM results show distinctly different evolutions in both EM and temperature during different periods of the flare gradual phase. Especially the plasma temperature remains high and steady even far beyond the main flare phase, suggesting that additional processes besides the conventional cooling should occur above the flare loop tops. We discuss the results in the following sections.

\section{Discussion} \label{sec:dis}

\subsection{Flare Loop Geometry and Doppler shifts}

 \begin{figure}[htbp]
	\centering
	\includegraphics[width=\textwidth]{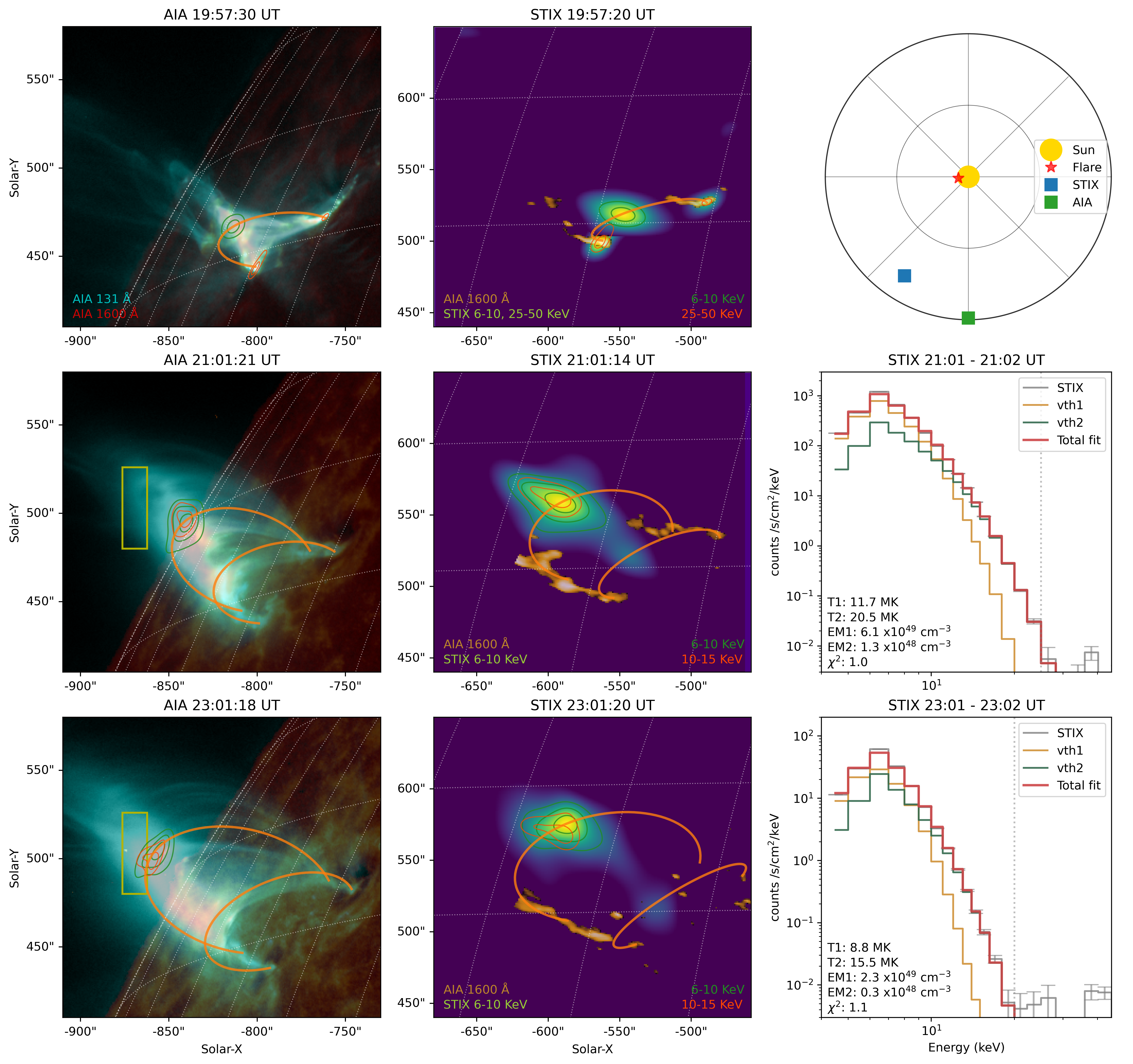}
	\caption{\small 3D flare loop geometry from multi-point observations by AIA and STIX. From top to bottom show observations during the flare impulsive phase, EGP, and LGP, respectively. From left to right are AIA 131~\AA\ and 1600~\AA\ blended images, STIX HXR images on top of the reprojected flare ribbons in AIA 1600~\AA, and STIX energy spectra. The flare location on the Sun and Solar Orbiter location relative to the Sun and Earth are shown in the top-right panel. The yellow box in the AIA image is the same with that in Figs.~\ref{fig:maps}\&\ref{fig:emt}. The green and orange contours in the left and middle panels show STIX HXR sources above 50\% of the peak emission in two energy bands. STIX loop-top sources are reprojected to the AIA's perspective and shifted to the modeled loop top. The semi-circular orange curves are representative 3D flare loops as seen in AIA and STIX. The STIX energy spectra are fitted with two isothermal functions, where the vertical dotted lines show maximum energies for fitting. The STIX observing time is corrected for the difference of light traveling to Solar Orbiter and Earth.}
	\label{fig:stix}
\end{figure}

In the flare, IRIS observes weak Doppler blue shifts in the spectral profile near the flare loop-top region, showing bulk plasma motions with a velocity component toward the observer. We examine the flare loop orientation to understand the direction of plasma motions. We use multi-point observations from SDO and Solar Orbiter \citep{Muller2020} to reconstruct the three-dimensional (3D) flare loops. Solar Orbiter was located at a heliocentric distance of 0.83~AU from the Sun at the time of flare, and the Extreme Ultraviolet Imager (EUI) did not take images of the solar disk but the Spectrometer Telescope for Imaging X-rays \citep[STIX][]{Krucker2020} recorded HXR emissions of the flare. Solar Orbiter was separated from the Earth by about $-33^\circ$ in longitude (see the upper-right panel in Fig.~\ref{fig:stix}), which provides an appropriate angle for using AIA and STIX to perform a 3D triangulation. We reconstruct STIX HXR images in different energy ranges using the MEM\_GE algorithm \citep{Massa2020,Massa2023}, which show two footpoint sources in 25 -- 50 keV in the flare impulsive phase and loop-top sources in lower energies during the whole flare (Fig.~\ref{fig:stix}). Since the distance of Solar Orbiter from the Sun is too far ($>$0.75~AU) to provide accurate pointing for STIX \citep{Warmuth2020}, we correct this offset issue by manually shifting the nonthermal footpoint sources onto the flare ribbons observed in AIA 1600~\AA\ during the impulsive phase. During the flare gradual phase, STIX detected HXR emissions only from relatively low energy ranges, $<$25~KeV before $\sim$22:00~UT and $<$15~keV afterward (Fig.~\ref{fig:timeline}a). We confirm these are mainly thermal emissions by fitting the HXR spectra with two isothermal functions (see the energy spectra in Fig.~\ref{fig:stix}). We obtain HXR images in 6 -- 10 and 10 -- 15 keV energy bands, which show thermal loop-top sources located between two semi-parallel flare ribbons.

We use AIA 131~\AA, 1600~\AA, and STIX 6 -- 10 keV images to reconstruct the flare loops in 3D (Fig.~\ref{fig:stix}). We use a semi-circular loop shape to place two footpoints on the AIA 1600~\AA\ ribbons, adjusting the loop to align with the 131~\AA\ flare loops and ensuring that the loop top is co-spatial with STIX sources. During the gradual phase of the flare, STIX imaged two loop-top sources, a primary one and a less-bright, extended source in the southwest (Fig.~\ref{fig:stix}). The two loop-top sources nicely correspond to the two brightest regions of hot emissions in AIA. Thus we reconstruct two loops that give a general representation of the 3D flare arcade.

The results show that the flare loops have a large elevation angle above the solar surface and tend to fall toward the observer. Additionally, the eastern half of the loops seen in AIA (where the IRIS raster is focused on) are farther away from the observer than the western half (Fig.~\ref{fig:stix}). The loop orientation suggests that the observed Doppler blue shifts are resulting from plasmas moving upward along the eastern half of the loop, which indicates chromospheric evaporation of hot plasmas. The observed velocities are relatively small ($<$20 km/s) along the line of sight, which can be explained in terms of a small tilt angle of the loop in the plane of sky as evidenced in the 3D loop geometry, as well as a measuring location close to the loop top rather than footpoints. Projecting the STIX loop-top sources to the Earth's view (left panels in Fig.~\ref{fig:stix}) shows that the IRIS slits are located above the loop-top source during EGP, and cover the loop-top region of HXR emissions later during LGP as the post-flare loops grow higher (e.g., $\sim$23~UT).

The Doppler velocities are observed to vary in different locations and at different times. In particular, higher velocities are found in the IRIS slits at higher altitudes above loop tops (green symbols in Fig.~\ref{fig:timeline}c), which suggests larger evaporating velocities probably from hot plasmas associated with newly reconnected field lines. This velocity-height association exists throughout the long-duration gradual phase, in agreement with the presence of on-going magnetic reconnection after the eruption. During EGP, the Doppler velocities tend to slightly decrease with time. This decrease can be a result of the continuous rising of post-flare loops in height, which makes the IRIS slits capture more lower-altitude plasmas as time passes. When the strongest intensities are observed around 21:10 -- 22:00 UT, nearly zero Doppler shifts are detected (dark orange symbols in Fig.~\ref{fig:timeline}b,c), which suggests that the plasmas in dense loop tops are almost at rest. During LGP, some weak red shits are also detected (for example, around 23~UT), and the velocities become disorderly toward the end of the IRIS observations. These line shifts can be caused by plasma motions in different directions in the loop tops and cusps, which are evident in AIA 131~\AA\ images (Fig.~\ref{fig:maps} and its animation). The spectral emissions are too weak to resolve plasma motions in individual loops, and weak intensities during LGP also result in large uncertainties of spectral fitting especially for those at higher altitudes. The increasing velocities during LGP suggest large disturbances in flaring plasmas during the late phase.

\subsection{Nonthermal Motions in Supra-arcade and Loop-top Region}

IRIS spectral observations show excess line broadening with large nonthermal velocities in the flare supra-arcade and loop-top region throughout the long-duration gradual phase. In AIA 131~\AA\ channel that has dominated \ion{Fe}{21} contribution, different features and loop dynamics are imaged during different phases of the flare, which may contribute to the nonthermal motions of hot plasmas.

During EGP, the nonthermal velocities are up to $\sim$130~km/s, when IRIS slits captured the bottom of the supra-arcade fan/plasma sheet as evidenced in the AIA images (Fig~\ref{fig:maps}). Especially before $\sim$21~UT, IRIS slits are almost completely on the supra-arcade fan, where large nonthermal velocities ($\sim$50 -- $>$100 km/s) are detected and exhibit similar characteristics for all slit positions (Fig.~\ref{fig:timeline}d). The nonthermal line broadening implies the presence of turbulence in the current sheet region above the flare arcade, which was also reported in some other cases \citep{Ciaravella2008,Doschek2014,LiY2018,Warren2018,Shen2023}. The results for this flare are generally of the same magnitude as previous cases. We note that the nonthermal velocities decrease rapidly during 20:45 -- 21:00~UT, accompanied by a simultaneous sharp increase in line intensities (Fig.~\ref{fig:timeline}). This result suggests a transition from supra-arcade plasmas to dense flare loop tops as the loops rise and start to appear in the FOV of IRIS raster, which is also evidenced in AIA 131~\AA\ images. The nonthermal velocities then gradually decrease with time, which coincides with the continuous rising of post-flare loops in height as observed in AIA (Fig.~\ref{fig:slit}). Since the supra-arcade and loop-top emissions are superposed together along the line of sight, the rise of the post-flare loop system results in varying plasma components captured by the IRIS spectrograph. In particular, higher velocities are observed at higher altitudes, probably due to a larger fraction of supra-arcade plasma along the line of sight. The observations provide clear evidence for the presence of turbulence in the supra-arcade plasmas.

The post-flare loop system keeps growing and then fades during LGP. The IRIS slits are located on numerous loop cusps in front of flare loop tops (Fig.~\ref{fig:maps}). Strong nonthermal velocities are also detected in this region and last during the very late phase. These velocities do not evolve as smoothly as earlier during EGP but are more disordered, which can be caused by a superposition of plasma flows moving in different directions. Indeed, AIA images show a plethora of contracting flare loops and loop cusps appearing in the IRIS raster region, consisting multiple plasma flows in the numerous cusp tips and loop tops. In addition, the flare loop-top region itself involves complex dynamics and possible unresolved plasma motions, as a result of reconnection outflows impinging the flare loops below \citep[see, for example,][]{Reeves2020,ChenB2024}. The spectral profiles, which we integrate over the loop-top region, still show a Gaussian shape, and the spectral emission is too weak to resolve different plasma flows. For different slit locations, higher altitudes generally have larger velocities, similar to that during EGP. This height association, similar to the Doppler velocity distribution, can be a natural result of the fact that the reconnection site is located high above in the corona during the flare gradual phase. The nonthermal velocities are still high toward the flare end, indicating the existence of strong nonthermal motions above the flare loops even $\sim$4 hours after the peak of the flare.

\subsection{Plasma Heating and Cooling}

DEM analysis shows that the supra-arcade plasmas in the region close to loop tops have the highest temperature during the flare (Fig.~\ref{fig:emt}). The temperature is about 18~MK ($\log T$ = 7.25) after the flare SXR peak, even hotter than the plasma sheet at higher altitudes ($\log T$ = 7.1; see PS\#3 in \citealt{Gou2024}). Moreover, during the long-duration gradual phase of the flare, the plasma temperature stays high in this region, still as high as $\sim$10~MK even about 4 hours after the flare peak (Fig.~\ref{fig:emt}). The observations show that the plasma temperature does not cool down following normal conductive and radiative cooling processes, which are in a timescale of several to a few tens of minutes for typical flare plasmas \citep[e.g.,][]{Cargill1995}. Thus there should be suppressed cooling or additional heating occurring in the region above flare loops.

In this flare, the IRIS slits are located right on the region with highest temperatures (Fig.~\ref{fig:emt}), which provides spectroscopic insights into the hot plasma. By focusing on the same region, we find that the temperature decreases from $\sim$18~MK to 10~MK with almost constant EM during EGP (Fig.~\ref{fig:dem}), showing inevitable cooling of hot plasmas probably due to dominant conductive cooling in the beginning. While the temperature stays steady at around 10~MK with decreasing EM during LGP, when significant radiative cooling should also occur and take over conductive cooling eventually. IRIS observed large nonthermal velocities of hot plasmas in the same region throughout the gradual phase, and the strong nonthermal motions, by the turbulence or numerous unresolved plasma flows, could contribute to the slowed-down cooling. On one hand, the plasma turbulence can suppress thermal conduction \citep{Allred2022,Xie2023}. On the other hand, downflows and retracting loops can contribute to local heating \citep{Reeves2017,Longcope2018}. AIA observations also demonstrate a lot of downflows and shrinking loops/cusps in the region above flare loops, even in the very end. Particularly there are still strong nonthermal plamsa motions toward the end of the prolonged gradual phase, which should contribute to the steady, high temperature in the same region. It is difficult to know from these observations how much the microscopic nonthermal motions contribute to the plasma heating/cooling or whether there are some other responsible processes. Flare loop modeling and theoretical approaches are helpful for further investigation.

\section{Summary and Conclusion} \label{sec:summary}

We study the long-duration gradual phase of an eruptive X2.3 flare on 2023 February 17, focusing on the flare loop-top region where IRIS \ion{Fe}{21} spectral emissions were observed. We examine the evolution of spectral parameters from IRIS and thermodynamic evolution of flare loops in AIA. We summarize our findings below.
\begin{enumerate}[noitemsep]
    \item IRIS has continuously observed \ion{Fe}{21} spectral emissions from hot plasmas around the loop-top region for about 4 hours, covering the whole decay of the X-class flare. Continuous growth and then fade of the post-flare arcade make the spectral emissions vary in both time and space.
    \item Small Doppler blue shifts ($<$20~km/s) are detected, indicative of bulk plasma motions from chromospheric evaporation of hot plasma. Excess line broadenings with significant nonthermal velocities (up to 130~km/s) are present throughout the flare gradual phase.
    \item Large nonthermal velocities are detected in the bottom of the supra-arcade fan/plasma sheet, suggestive of the presence of turbulence in the flare current sheet region. 
    \item Disorganized nonthermal velocities are present above flare loops during the late flare phase, where numerous contracting loop cusps and loop tops are observed, indicative of irregular nonthermal motions of unresolved plasma flows in this region.
    \item Larger Doppler shifts and higher nonthermal velocities are found at higher altitudes above the flare loops, suggesting that stronger bulk plasma motions and stronger turbulent/nonthermal motions are associated with newly reconnected field lines, in the context of on-going magnetic reconnection during the flare gradual phase.
    \item The plasma temperature above the loop top is the highest during the flare, and it stays high and steady even far beyond the main flare phase. The observed large nonthermal motions in the same region may contribute to the suppression of cooling or additional heating above flare loops.
\end{enumerate}

The observations provide spectroscopic insights into the hot flaring plasma, particularly through a whole life of this long-duration flare event. The results reveal complex plasma dynamics associated with nonthermal plasma motions above the flare loops, and demonstrate the power of high-resolution imaging and spectroscopic observations for understanding the flare dynamics. The upcoming Multi-slit Solar Explorer \citep[MUSE;][]{DePontieu2020,Cheung2022}, with a 35-slit EUV spectrograph and an AR-scale FOV to capture the entire flare loop system, will provide unprecedented spectroscopic observations to understand the detailed plasma dynamics occurring during solar flares.

\vspace{20pt}
\begin{acknowledgments}
    This work is supported by contract 8100002705 from Lockheed-Martin to SAO. IRIS is a NASA small explorer mission developed and operated by LMSAL with mission operations executed at NASA Ames Research Center and major contributions to downlink communications funded by ESA and the Norwegian Space Centre.
\end{acknowledgments}

\bibliography{ref}{}
\bibliographystyle{aasjournal}

\end{document}